\documentclass[fleqn,usenatbib]{mnras}

\usepackage[T1]{fontenc}
\usepackage{ae,aecompl}
\usepackage{graphicx}	% Including figure files
\usepackage{amsmath}	% Advanced maths commands
\usepackage{amssymb}	% Extra maths symbols
\usepackage{txfonts}
\usepackage{natbib,longtable,times,multicol}
\usepackage{caption}
\usepackage[caption = false]{subfig}
\usepackage{verbatim}
\usepackage{makecell}
\usepackage{epstopdf}

\title[Relativistic reflection in AGN at z=0.5--4]{Relativistic reflection from accretion disks in the population of Active Galactic Nuclei at z=0.5--4}

\author[]{
 \parbox[t]{18cm}{L. Baronchelli$^1$\thanks{E-mail: blinda@mpe.mpg.de}, K. Nandra$^1$, J. Buchner$^2$}\\\\
$^{1}$Max-Planck-Institut f\"{u}r extraterrestrische Physik, Giessenbachstrasse 1, 85748 Garching bei M\"{u}nchen, Germany\\
$^{2}$Instituto de Astrofisica, Facultad de Fisica, Pontificia Universidad Catolica de Chile, Casilla 306, Santiago 22, Chile\\
}

\date{Accepted 2018 July 25. Received 2018 July 25; in original form 2018 April 16}

\pubyear{2018}

\begin{document}
\label{firstpage}
\pagerange{\pageref{firstpage}--\pageref{lastpage}}
\maketitle

\begin{abstract}

We report the detection of relativistically broadened iron K$\alpha$ emission in the X-ray spectra of AGN detected in the 4Ms CDF-S.  
Using the Bayesian X-ray analysis (BXA) package, we fit 199 hard band (2--7 keV) selected sources in the redshift range \textit{z}=0.5--4 with three models: (i) an absorbed power-law, (ii) the first model plus a narrow reflection component, and (iii) the second model with an additional relativistic broadened reflection. The Bayesian evidence for the full sample of sources selects the model with the additional broad component as being 10$^{5}$ times more probable to describe the data better than the second model. For the two brightest sources in our sample, CID 190 (\textit{z}=0.734) and CID 104 (\textit{z}=0.543), BXA reveals the relativistic signatures in the individual spectra. We estimate the fraction of sources containing a broad component to be 54$^{+35}_{-37}\%$ (107/199 sources). Considering that the low signal-to-noise ratio of some spectra prevents the detection of the broad iron K$\alpha$ line, we infer an intrinsic fraction with broad emission of around two thirds. The detection of relativistic signatures in the X-ray spectra of these sources suggests that they are powered by a radiatively efficient accretion disk. Preliminary evidence is found that the spin of the black hole is high, with a maximally spinning Kerr BH model (\textit{a}=1) providing a significantly better fit than a Schwarzschild model (\textit{a}=0). Our analysis demonstrate the potential of X-ray spectroscopy to measure this key parameter in typical SMBH systems at the peak of BH growth.

\end{abstract}


%%%%%%%%%%%%%%%%%%%%%%%%%%%%%%%%%%%%%%%%%%%%%%%%%%

%%%%%%%%%%%%%%%%% BODY OF PAPER %%%%%%%%%%%%%%%%%%

\section{Introduction}
\label{sec:intro}

The X-ray emission from active galactic nuclei (AGN) is believed to arise when the optical/UV photons radiated from the accretion disk are inverse Compton scattered by a corona of hot electrons surrounding the supermassive black hole (SMBH) \citep[e.g.,][]{Sunyaev1980,Haardt1991,Haardt1993,Zdziarski1998,Jovanovic2009}. 
The primary X-ray photons illuminate and are reprocessed by material surrounding the nucleus, including the accretion disk and the obscuring torus, producing a co-called Compton reflection spectrum \citep[e.g.,][]{George1991}. Thus, the reflection component of the AGN X-ray spectrum contains information about the geometrical structure and the dynamics of the matter surrounding the SMBH \citep{Reynolds1999}. The most prominent features of the reflection component are the iron (Fe) K$\alpha$ feature at 6.4 keV in the rest frame and the Compton hump peaking at energies around 30 -- 40 keV \citep[e.g.][]{Pounds1990,Nandra1994}. The Compton hump is only produced when the matter surrounding the SMBH is Compton thick ($\mathrm{N_H \gtrapprox 10^{24}\; atoms\; cm^{-2}}$ which is approximately the inverse Thomson optical depth $\mathrm{\tau_{Thom}}$) while the Fe K$\alpha$ line can be produced also when the reflecting material is Compton-thin \citep{Lightman1988,Krolik1999,Ricci2014}. Narrow components to the Fe K$\alpha$ emission are seen nearly ubiquitously in the spectrum of nearby AGN \citep{Yaqoob2004,Nandra2007,Ricci2014} and generally attributed to an origin in the molecular torus \citep{Krolik1987,Nandra2006}. However, when the reflection component arises from the innermost region of the accretion flow such as the accretion disk, the gravitational field of the SMBH affects the shape of the line \citep{Fabian1989}. The resulting line profile is broadened and skewed by light bending, gravitational redshift and relativistic Doppler shifts and in extreme cases the line can extend in energy from $\sim$3--7 keV \citep{Fabian2000,Lee2002,Yaqoob2007}, indicating an origin close to the innermost stable orbit of the accretion disk. Because this depends on the black hole spin, the study of broadened Fe K$\alpha$ lines represents an important probe of general relativistic effects and the dynamics of the SMBH, as well as the matter immediately surrounding it \citep{Yaqoob2002,Reynolds2003,Fabian2005}.

The clearest evidence for a relativistic broadened Fe K$\alpha$ line was found by \citet{Tanaka1995} in the spectrum of the nearby Seyfert galaxy MCG-6-30-15 observed by \textit{ASCA}. Since then, the presence of a skewed and broadened Fe K$\alpha$ line profile has been confirmed in the X-ray spectra of many other bright, nearby AGN \citep{Nandra1997,Guainazzi2006,Nandra2007,CallePerez2010}. While the relativistic phenomena therefore seem to be reasonably widespread in nearby SMBH systems, it is much more difficult to establish whether they are common in typical AGN at higher redshifts, and specifically those responsible for the bulk of the accretion power in the Universe, during the peak of black hole activity at z=0.5-4. 

Past work aiming to do this has used stacking of the Fe K$\alpha$ line in large samples of observations with low signal to noise ratio \citep{Streblyanska2005,Chaudhary2012}. In principle, this allows one to infer the global properties of the population in cases where it would be uninformative to fit single sources individually. For example, by analyzing the mean rest-frame spectra of a sample of type-1 and type-2 AGN over a broad redshift range in the Lockmann hole from the \textit{XMM-Newton}, \citet{Streblyanska2005} presented evidence for broad Fe K$\alpha$ line emission.
However, it has been found that stacking the spectra can introduce artificial broadening in the Fe K$\alpha$ line, especially in samples where the sources have a wide redshift distribution \citep{Chaudhary2012}. \citet{Chaudhary2012} nonetheless concluded after performing rest-frame stacking of a sample of 248 AGN from the 2XMM catalog that the average Fe K$\alpha$ line profile is best represented by a combination of narrow and broad line. On the other hand, \citet{Corral2008} computed the averaged rest-frame spectrum of 600 \textit{XMM-Newton} observation of type-1 AGN without finding compelling evidence for any significant broad line component.

Evidence for broad Fe K$\alpha$ line emission in stacking studies has also been reported in \citet{Falocco2013}, where the XMM CDF-S spectra were averaged, and in \citet{Falocco2014},  where the authors explored the spectra of an AGN sample built using the 2nd XMM serendipitous survey and the VCV catalog \citep{Veron2006,Veron2010}. The detection in \citet{Falocco2014} is compatible with the upper limit obtained in \citet{Corral2008,Corral2011}. \citet{Liu2016} also found that the Fe K$\alpha$ might get broader by higher Eddington ratio. \citet{Falocco2012,Falocco2013,Falocco2014,Liu2016} support the spectral stacking with simulations that quantify the broadening introduced by the stacking itself. Hence, these studies are free of artificial broadening.

Thus, the ubiquity of the relativistic broadened Fe K$\alpha$ line in AGN X-ray spectra outside the local Universe is still controversial.

In this work, we aim to investigate the presence of the relativistic broadening of the Fe K$\alpha$ in the spectra of the AGN population as observed with Chandra using a different approach, specifically by using the Bayesian X-ray analysis (BXA) spectral fitting method \citep{Buchner2014}. This allows information to be extracted from the individual spectra even if of low signal-to-noise ratio, and the results combined to make inferences about the population as a whole. In Sect. \ref{sec:method} we define the method and the spectral modeling applied to our data, including the definition of the sample selection. The results of the analysis are reported in Sect. \ref{sec:res} and discussed in Sect. \ref{sec:Summary_Conclusion}.

Throughout this work, we adopt $\Omega_m$= 0.272, $\Omega_\Lambda$= 0.728, and $H_0$ = 70.4 km s$^{-1}$ Mpc$^{-1}$ \citep{Komatsu2011}.  Errors are quoted at the 90\% confidence level unless otherwise specified.   

\section{Method}
\label{sec:method}

\subsection{Data and Sample selection}
\label{sec:samp}
In this work we wish to explore the X-ray spectral properties of sources at the peak of supermassive black hole activity. Our sample is chosen from the 4Ms exposure of the \textit{Chandra} Deep Field-South (CDF-S) \citep{Xue2011}, totaling 51 observations over an area of 464.5 arcmin$^2$. This deep \textit{Chandra} exposure gives us the opportunity of studying the Fe K$\alpha$ line of AGN spectra up to redshifts of $z\sim 4$.

The source and background spectra employed in this work are a selected subsample of the spectra extracted by \citet{Brightman2014} analyzed by \citet{Buchner2014} from the CDF-S 4Ms, which is based on the source catalog of \citet{Rangel2013} (hereafter R13), using data reduction methods following the work of \citet{Laird2009}. The background extraction region around any given source is constructed so that it contains at least 100 counts after masking all the other sources out. Our sources are hard band (2--7 keV) selected and as in \citet{Rangel2013} are considered significant if the Poisson probability that the observed counts are a background fluctuation is less than $\mathrm{4\times 10^{-6}}$. We used spectroscopic (preferably) or photometric redshifts from the work of \citet{Hsu2014}, following \citet{Buchner2014}. 
In addiction to this, we restrict the analysis to sources with redshift lower than 4, both due to limitations in models and because above this redshift the sources are few and of low signal-to-noise ratio. 
To ensure the reliability of the fits to the Fe K $\alpha$ feature, we exclude from the analysis sources with less than 20 counts in the 4 -- 7 keV energy range. 
The selected sample is composed of 123 sources with spectroscopic redshift and 76 sources with photometric redshift for a total of 199 sources. 
Note that in the case of sources with photometric redshift, the probability distribution function is used in the spectral fitting rather than a single value, thus accounting for the systematic uncertainties in the photo-z determination \citep{Buchner2015}. 
In total, the selected sources show 30,667 counts in the 4--7 keV rest frame energy range. The total number of counts in this energy range in our sample is comparable to bright sources in the local Universe. \citet{Nandra2007}, for example, imposed a lower limit of 30,000 counts in the 2-10 keV band when constructing their sample. Once combined, our data quality should be sufficient to provide constraints on the broad iron line  \citep[see e.g.][]{Guainazzi2006,Mantovani2014}.

\subsection{Model comparison overview}

We employ the Bayesian X-ray Analysis \citep[BXA,][]{Buchner2014} software to fit the X-ray spectra and compare models. BXA is a Bayesian framework to determine the best-fitting models and their parameter constraints for X-ray spectra. It can also be used as a robust statistical tool to determine if the data are better fit by a model containing a relativistically blurred component or by a simpler model. A key features is that it allows us to analyze a large sample of low signal to noise ratio observations and make inferences for the population without the need to stack the spectra. BXA computes for each model the Bayesian evidence Z, which is the likelihood integrated over the parameter space (see Section \ref{sec:modcomp}). The ratio $\mathrm{Z_{M_1}/Z_{M_2}}$ can then be used to compare the models $\mathrm{M_1}$ and $\mathrm{M_2}$.

We compared three models to the X-ray data. We first fit the spectra using a simple absorbed power-law to model the intrinsic X-ray emission. We then added a narrow reflection component to the intrinsic continuum, presumed to arise from distant material such as the molecular torus. Finally we added a blurred reflection component attributed to the accretion disk. The primary continuum of the AGN X-ray spectra was modeled with a redshifted power-law \texttt{zpowerlw}.
The narrow reflection was modeled with the \texttt{pexmon} model \citep{Nandra2007}, which combines a reflected power-law in neutral medium \citep{Magdziarz1995} with self consistently generated Fe K$\alpha$ (6.4 keV), Fe K$\beta$ (7.059 keV), Ni K$\alpha$ (7.48 keV) and the Fe K$\alpha$ Compton shoulder. We modeled the relativistic broadened component by convolving another \texttt{pexmon} component with a kernel representing different models of a relativistic accretion disk around the SMBH. \begin{comment}(Figure \ref{fig:models})\end{comment}
We consider both the case of a Schwarzschild SMBH, with spin a = 0, and the case for a Kerr SMBH, with a = 1, given that the shape of the emission line depends on the metric used to describe the space-time of the SMBH.
The first is achieved by convolving the \texttt{pexmon} with the \texttt{rdblur} model \citep[a=0, ][]{Fabian1989} and the second the \texttt{kdblur} model \citep[a=1, ][]{Laor1991}. \begin{comment}(Figure \ref{fig:Iron})\end{comment} 
Convolving the \texttt{pexmon} with these relativistic kernels has the advantage that the entire reflection components is blurred and not only the Fe K$\alpha$ line.

\begin{center}
\begin{figure}
\centering
\includegraphics[scale=0.36]{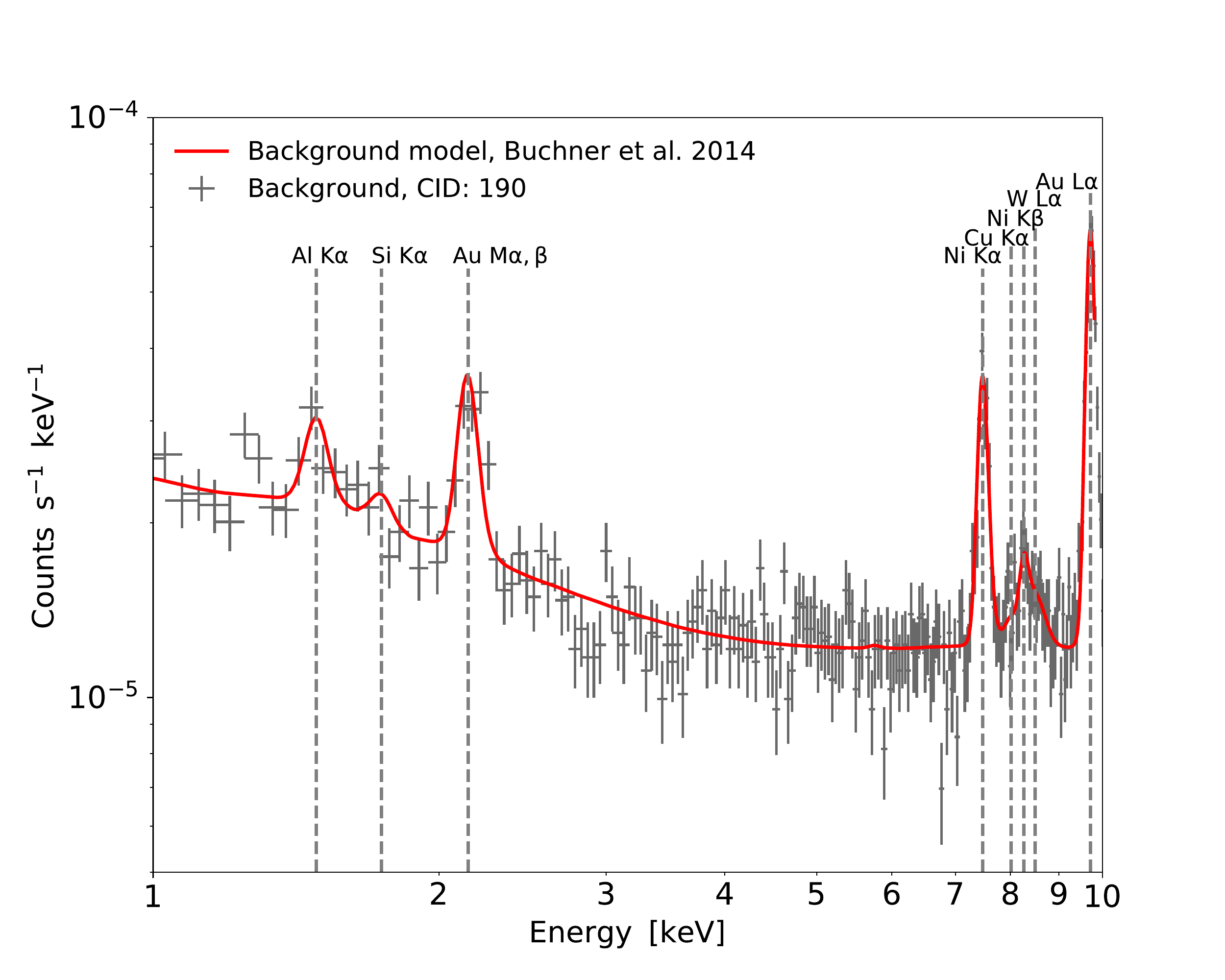}
\caption{Example of the fitted background for the source CID 190. The background model \citep{Buchner2014} is composed by eight narrow Gaussians describing the particle emission in the detector and a continuum component, described by a \texttt{box1d} \textsc{Sherpa} model \citep{Bartalucci2014}. In addition to the eight narrow Gaussians, the continuum also includes two broad Gaussians at energies lower than 1keV. The background model is first fitted to the background spectrum of each individual source. The best fit background parameters so-determined are subsequently frozen in the fit to the source spectrum, which is not background subtracted.}
\label{fig:background}
\end{figure}
\end{center}

The free and frozen parameters of the three models are listed in Table \ref{powerlw}, \ref{pexmon} and \ref{blur} as well as the intervals chosen as the priors for BXA.
We obtain the likelihood and the best fit parameters using the C-statistic \citep{Cash1979}. This statistic is appropriate with low signal to noise data where the counts are sampled from the Poisson distribution. When applying the C-statistic we can not subtract the background, so we model this simultaneously with the source spectrum. We use the Chandra background model from \citet{Buchner2014} (see Figure \ref{fig:background}). 

To be able to use the \textit{Chandra} background (Figure \ref{fig:background}) implementation from \citet{Buchner2014} we use the \textsc{sherpa} \citep{Freeman2001} implementation of BXA (\textsc{ciao} version 4.8). However, the two convolution models \texttt{rdblur} and \texttt{kdblur} used to blur the \texttt{pexmon} are available only in \textsc{xspec} \citep{Arnaud1996}. We therefore produced a table model of the convolved \texttt{pexmon}, \texttt{kdblur(rdblur)} and \texttt{kdblur(pexmon)}, from \textsc{xspec} (version 12.9.0) and imported the two table models into \textsc{sherpa}. This procedure also has the advantage of reducing the fitting time required by BXA. To further speed the fitting procedure, we used \textsc{xspec} and the BXA module \texttt{RebinnedModel} to approximate the narrow \texttt{pexmon} component in models \ref{eq:pexmon} and \ref{eq:blur} by interpolating it over a smaller subsample of its prior space. 

\begin{comment}
\begin{center}
\begin{figure}
\centering
\includegraphics[scale=0.34]{models-eps-converted-to.pdf}
\caption{.}
\label{fig:models}
\end{figure}
\end{center}

\begin{center}
   \begin{figure}
     \subfloat{%
       \includegraphics[width=0.45\textwidth]{Iron_rdblur-eps-converted-to.pdf}
     }
     \\
     \subfloat{%
       \includegraphics[width=0.45\textwidth]{Iron_kdblur-eps-converted-to.pdf}
     }
     \caption{Comparison of Fe K$\alpha$ lines at different disk inclination angles (5$^{\circ}$, 30$^{\circ}$ and 75$^{\circ}$). The emission line becomes broader and more skewed for higher disk inclination angles. The \textit{left} figure shows the model (\texttt{rdblur}) for Schwarzschild BHs with spin a = 0. The figure on the \textit{right} shows the model (\texttt{kdblur}) for Kerr BHs with spin a $\sim$ 1. When the \texttt{rdblur} at low inclination angles is used to convolve the \texttt{pexmon} it is still possible to distinguish the Fe K$\beta$ (7.059 keV) and the Ni K$\alpha$ (7.48 keV) lines.}
     \label{fig:Iron}
   \end{figure}
\end{center}
\end{comment}

%\section{Models}
\subsection{Model parameters}
\label{sec:models}
To define the models in a Bayesian framework we need to describe the prior distributions (or priors) of the parameters. The priors are the information about a parameter $\theta$ that is combined with the probability distribution of the data to generate the posterior distribution \citep{Gelm2002}. In order to chose an appropriate prior distribution we need to take into account the information that the priors are going to contain and the properties of the resulting posterior distribution. We constrain the priors of our parameters to be limited to physical motivated intervals. However, we maintain the priors are as uninformative as possible to avoid biases in the posterior distribution. Note also that the prior choices for parameters that the models have in common are unimportant. However, the prior spaces of the parameters not shared by all the models have to be chosen more carefully.

We report in Table \ref{powerlw}, \ref{pexmon} and \ref{blur} the parameter space of the three models \ref{eq:powerlw}, \ref{eq:pexmon} and \ref{eq:blur} respectively. If the parameter is left free to vary we show the range of accepted values. We choose log-uniform priors on the normalization and column density $\mathrm{n_H}$, and uniform priors on photon index and inclination (model \ref{eq:blur}).

The normalizations of the \texttt{pexmon} components (narrow and blurred) are modeled relative to the intrinsic power-law and may vary between 10$^{-2}$ and 10. The lower limit for the relative strength of the reflection is chosen on the basis that weaker components would imply an extremely small solid angle subtended by the accretion disk at the X-ray source, and in any event would be undetectable in the data. For the simplest scenarios, it is not expected that the relative normalization should exceed unity. This can be the case, however, if there is dramatic variability of the X-ray continuum, or if strong gravitational light bending is at play \citep{Miniutti2003} motivating the choice of our upper limit of $R=10$. 

Restricting the strength of the reflection components relative to the normalization of the primary power-law within reasonable physical limits reduces the prior volumes of the complex models.

\begin{table}
\caption{Parameter description for model \label{eq:powerlw} \texttt{zwabs*(zpowerlw)}. The model has three free parameters. \label{tab:params}}
\begin{tabular}{lllllll}
\hline \hline
 Comp.$^{a}$ & \multicolumn{1}{l}{No.$^{b}$} & Name$^{c}$ & Min & Max & Fixed val. & free \\ 
 \hline
zwabs & 1 & log(nH) & 20 & 26 & - & yes \\ 
 & 2 & Redshift & - & - & \multicolumn{1}{l}{z} & - \\ 
zpowerlw & 3 & PhoIndex & \multicolumn{1}{l}{1.1} & \multicolumn{1}{l}{2.5} & - & yes \\ 
 & 4 & Redshift & - & - & \multicolumn{1}{l}{z} &  -\\ 
 & 5 & $\log \mathrm{A_{pow}}$ & -10 & 1 & - & yes \\ 
 \hline
 \hline
\end{tabular}
\label{powerlw}
\flushleft 
\footnotesize{$^{a}$ Model component.}\\
\footnotesize{$^{b}$ Parameter number.}\\
\footnotesize{$^{c}$ Parameter name.}\\

\end{table}

\begin{table}
\caption{Parameter description for model \label{eq:pexmon} \texttt{zwabs*(zpowerlw + pexmon)}. The model has four free parameters: the column density $\mathrm{N_H}$, the photon index and the two normalizations. The strength of the unblurred reflection component $R_{pex}$ is measured relative to the power-law and is defined as the ratio of the normalization of the pexmon component ($A_{pex}$) to that of the power-law ($A_{pow}$).}
\begin{tabular}{lllllll}
\hline \hline
Comp.$^{a}$ & \multicolumn{1}{l}{No.$^{b}$} & Name$^{c}$ & Min & Max & Fixed val. & Free \\ \hline
zwabs & 1 & log(nH) & 20 & 26 & - & yes \\ 
 & 2 & Redshift & - & - & \multicolumn{1}{l}{z} & - \\ 
zpowerlw & 3 & PhoIndex & \multicolumn{1}{l}{1.1} & \multicolumn{1}{r}{2.5} & - & yes \\ 
 & 4 & Redshift & - & - & \multicolumn{1}{l}{z} & - \\ 
 & 5 & $\log \mathrm{A_{pow}}$ & -10 & 1 & - & yes \\ 
pexmon & 6 & PhoIndex & - & - & link to 3 & - \\ 
 & 7 & foldE & - & - & 800 & - \\ 
 & 8 & rel$\_$refl & - & - & \multicolumn{1}{l}{-1} & - \\ 
 & 9 & redshift & - & - & \multicolumn{1}{l}{z} & - \\ 
 & 10 & abund & - & - & \multicolumn{1}{l}{1} & - \\ 
 & 11 & Fe$\_$abund & - & - & \multicolumn{1}{l}{1} & - \\ 
 & 12 & Incl & - & - & \multicolumn{1}{l}{60} & - \\ 
 & 13 & $\log \mathrm{R_{pex}}$ & -2 & 1 & $\log \mathrm{\frac{A_{pex}}{A_{pow}}}$ & yes \\ \hline \hline
\end{tabular}
\label{pexmon}
\flushleft 
\footnotesize{$^{a}$Model component.}\\
\footnotesize{$^{b}$Parameter number.}\\
\footnotesize{$^{c}$ Parameter name.}\\

\end{table}

\begin{table}
\caption{Parameter description for model \texttt{zwabs*(zpowerlw + pexmon + blur(pexmon))} \label{eq:blur}. The model component blur could be \texttt{rdblur} or \texttt{kdblur}. The model has six free parameters: the column density $\mathrm{N_H}$, the photon index, the inclination of the broad component and the three norms. The strength of the blurred reflection component component $R_{blur}$ is measured relative to the power law and is defined as the ratio of the normalization of the blurred pexmon component ($A_{blur}$) to that of the power-law ($A_{pow}$).}
\begin{tabular}{lllllll}
\hline \hline
Comp.$^{a}$ & \multicolumn{1}{l}{No.$^{b}$} & Name$^{c}$ & Min & Max & Fix val. & Free \\ \hline
zwabs & 1 & log(nH) & 20 & 26 & - & yes \\ 
 & 2 & Redshift & - & - & \multicolumn{1}{l}{z} & - \\ 
zpowerlw & 3 & PhoIndex & \multicolumn{1}{l}{1.1} & \multicolumn{1}{l}{2.5} & - & yes \\ 
 & 4 & Redshift & - & - & \multicolumn{1}{l}{z} & - \\ 
 & 5 & $\log \mathrm{A_{pow}}$ & -10 & 1 & - & yes \\ 
pexmon & 6 & PhoIndex & - & - & link to 3 &  -\\ 
 & 7 & foldE & -& - & 800 & - \\ 
 & 8 & rel$\_$refl & - & - & \multicolumn{1}{l}{-1} & - \\ 
 & 9 & redshift & - & - & \multicolumn{1}{l}{z} & - \\ 
 & 10 & abund & - & - & \multicolumn{1}{l}{1} & - \\ 
 & 11 & Fe$\_$abund & - & - & \multicolumn{1}{l}{1} & - \\ 
 & 12 & Incl & - & - & \multicolumn{1}{l}{60} & - \\ 
 & 13 & $\log \mathrm{R_{pex}}$ & -2 & 1 & $\log \mathrm{\frac{A_{pex}}{A_{pow}}}$ & yes \\ 
\makecell{kdblur \\rdblur} & 14 & \makecell{Index \\ Betor10} & - & - & \multicolumn{1}{l}{\makecell{3 \\-2}} & - \\ 
 & 15 & Rin & - & - & 6 & - \\ 
 & 16 & Rout & - & - & 100/1000 & - \\ 
 & 17 & Incl & 10 & 85 & - & yes \\ 
pexmon & 18 & PhoIndex & - & - & link to 3 & - \\ 
 & 19 & foldE & - & - & 800 & - \\ 
 & 20 & rel$\_$refl & - & - & -1 & - \\ 
 & 21 & redshift & - & - & \multicolumn{1}{l}{z} & - \\ 
 & 22 & abund & - & - & 1 & - \\ 
 & 23 & Fe$\_$abund & - & - & 1 & - \\ 
 & 24 & Incl & - & - & link to 17 & - \\ 
 & 25 & $\log\mathrm{R_{blur}}$ & -2 & 1  & $\log \mathrm{\frac{A_{blur}}{A_{pow}}}$ & yes \\ \hline \hline
\end{tabular}
\label{blur}

\flushleft 
\footnotesize{$^{a}$Model component.}\\
\footnotesize{$^{b}$Parameter number.}\\
\footnotesize{$^{c}$Parameter name.}\\
\end{table}

\subsection{Model comparison in practice}
\label{sec:modcomp}
BXA calculates the Bayesian evidence (Z, also said marginal likelihood) using the Multimodal Nested Sampling Algorithm \citep[\texttt{MultiNest},][]{Skilling2004,Feroz2008,Feroz2009,Feroz2013} through its python wrapper \texttt{PyMultiNest} \citep{Buchner2014}. We use the Bayesian evidence Z to apply a model comparison using the Bayes Factor (BF) (see Equation \ref{eq:BF}) to determine which of the three fitted models describes the data better. The BF is the ratio of the marginal likelihood Z of two competing models. Thus the BF method will select the model with highest Z as the one better describing the data. The Bayesian evidence Z is the likelihood integrated over the prior distribution, thus the models with a large prior volume are naturally penalized. In our case, the broad component model has more free parameters, hence it is penalized by the BF method. To determine $\mathrm{Z_{tot}}$ for the entire sample we add the log(Z) of the single sources.

The outputs of BXA are the Bayesian evidence for the fitted model, the parameters posterior probability distribution and the marginal likelihood Z. 

The Bayes factor $\mathrm{B_{12}}$ (BF) is the ratio of the Bayes evidence $\mathrm{Z_i = P(D|M_i)}$, where $\mathrm{i = 1,2}$, of two competing models $\mathrm{M_1}$ and $\mathrm{M_2}$

\begin{equation} \label{eq:BF}
B_{12} = \frac{Z_1}{Z_2}
\end{equation}

\citep{Jeffreys1939}. Occam's razor is naturally implemented in the Bayes evidence (see Equation \ref{eq:natural_Occam}) since it is defined as the likelihood ($L(\theta)$) integrated over the whole parameter space ($\theta$) weighted by the priors ($P(\theta|M)$):
\begin{equation} \label{eq:natural_Occam}
Z = P(D) = \int L(\theta)P(\theta|M) \mathrm{d}\theta \simeq P(\theta) \mathrm{\delta} \theta L(\theta) \simeq \frac{\mathrm{\delta} \theta}{\Delta \theta}L(\theta)\theta,
\end{equation}
where $M$ is the model, $\theta$ is the parameters vector and $\mathrm{\delta}\theta / \mathrm{\Delta} \theta$ is the Occam's factor (OF) \citep{Jefferys1992}. The OF is the ratio between the posterior accessible volume ($\mathrm{\delta}\theta$) and the prior accessible volume ($\mathrm{\Delta}\theta$) and it prevents data over-fitting by penalizing the BF of more complex models, i.e. the models with the largest prior volume.

The values of the BF method can be interpreted using the Jeffrey scale, which strengthens its verdict roughly every time that the logarithm of the BF ($\mathrm{log(B_{ij})}$) increases by one in logarithmic space \citep{Robert2009}.

\section{Results}
\label{sec:res}

\begin{center}

   \begin{figure*}
     \subfloat{%
       \includegraphics[width=0.5\textwidth]{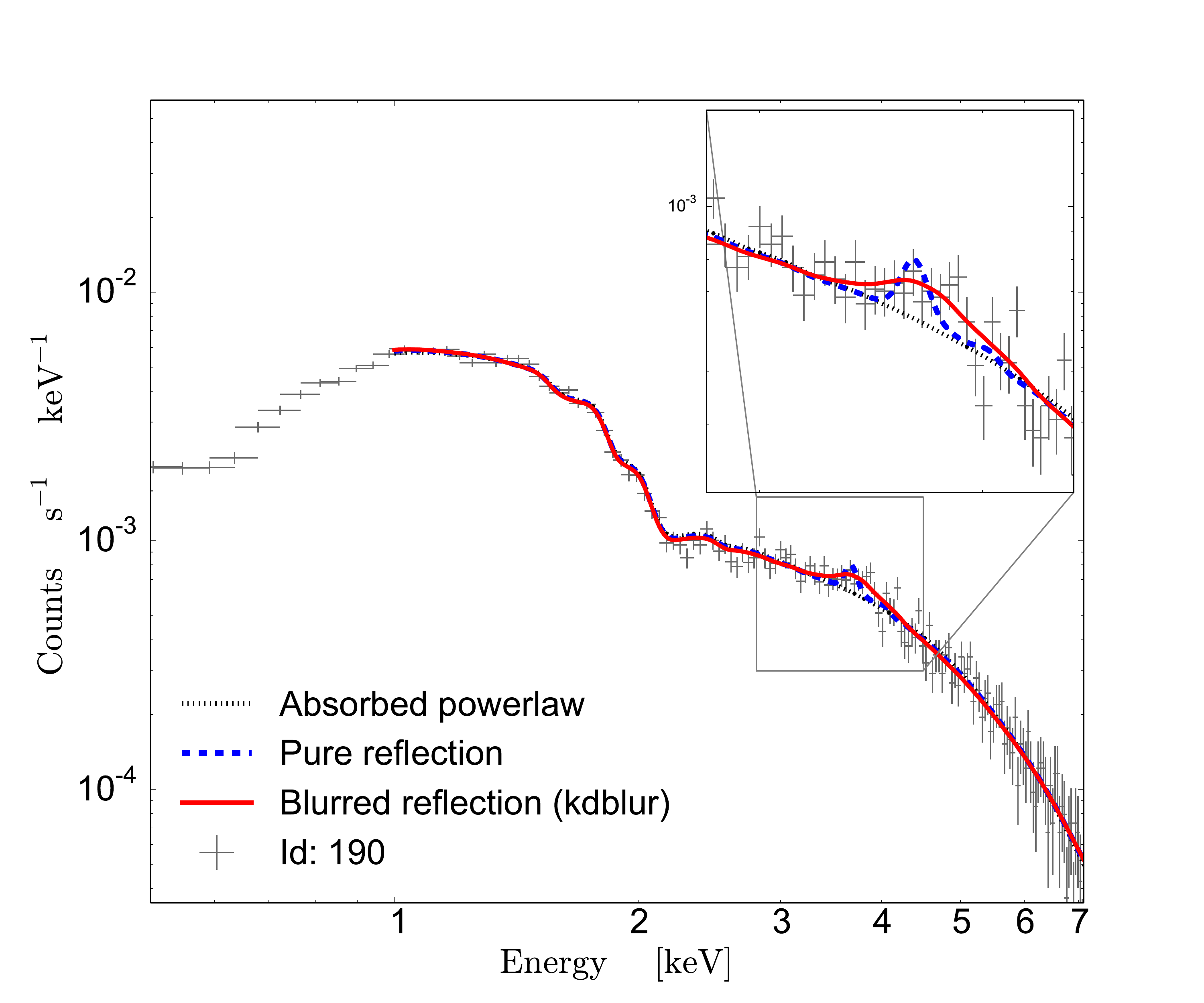}
     }
     \subfloat{%
       \includegraphics[width=0.5\textwidth]{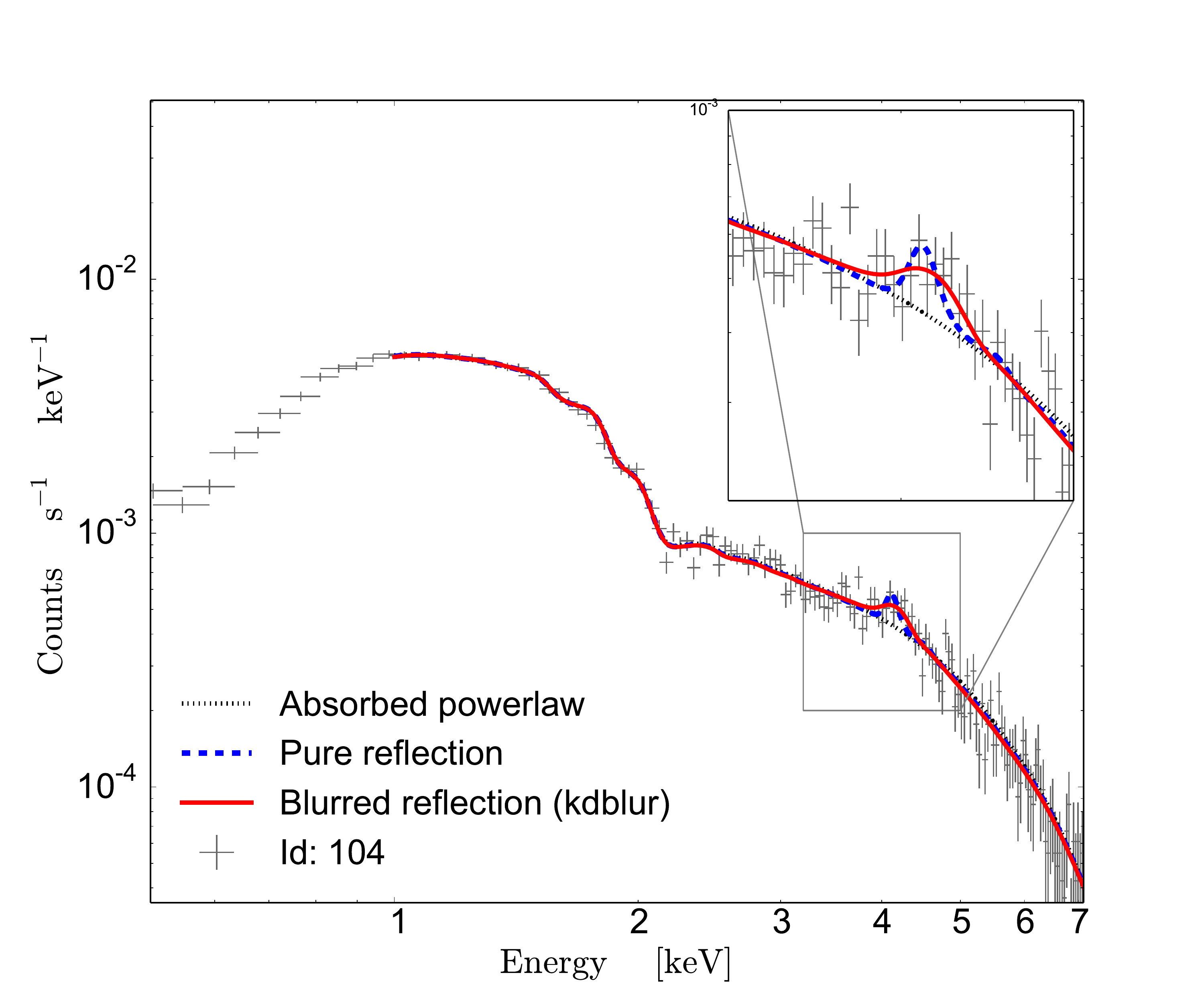}
     }
     \caption{The two brightest sources in the selected sample, CID 190 (\textit{Left}) and CID 104 (\textit{Right}) (Id number from R13) at spectroscopic redshift of 0.734 and 0.543 and with number of counts in the 4--7 keV of 2723 cts and 2502 cts respectively. CID 190 and CID 104 are the only two sources of the sample with number of counts in the 4--7 keV higher than 2000 cts. These two sources are selected by the BF method as better fitted by model \ref{eq:blur} with maximal spin. The models \ref{eq:powerlw}, \ref{eq:pexmon} and \ref{eq:blur} are shown in black (dotted), blue (dashed) and red (solid), respectively. Moreover, these sources are bright enough to constrain the inclination angle of the accretion disk modeled by the blurred component, albeit with large uncertainties.}
     \label{fig:weird}
   \end{figure*}
\end{center}
\begin{center}

   \begin{figure*}
     \subfloat{%
       \includegraphics[width=0.35\textwidth]{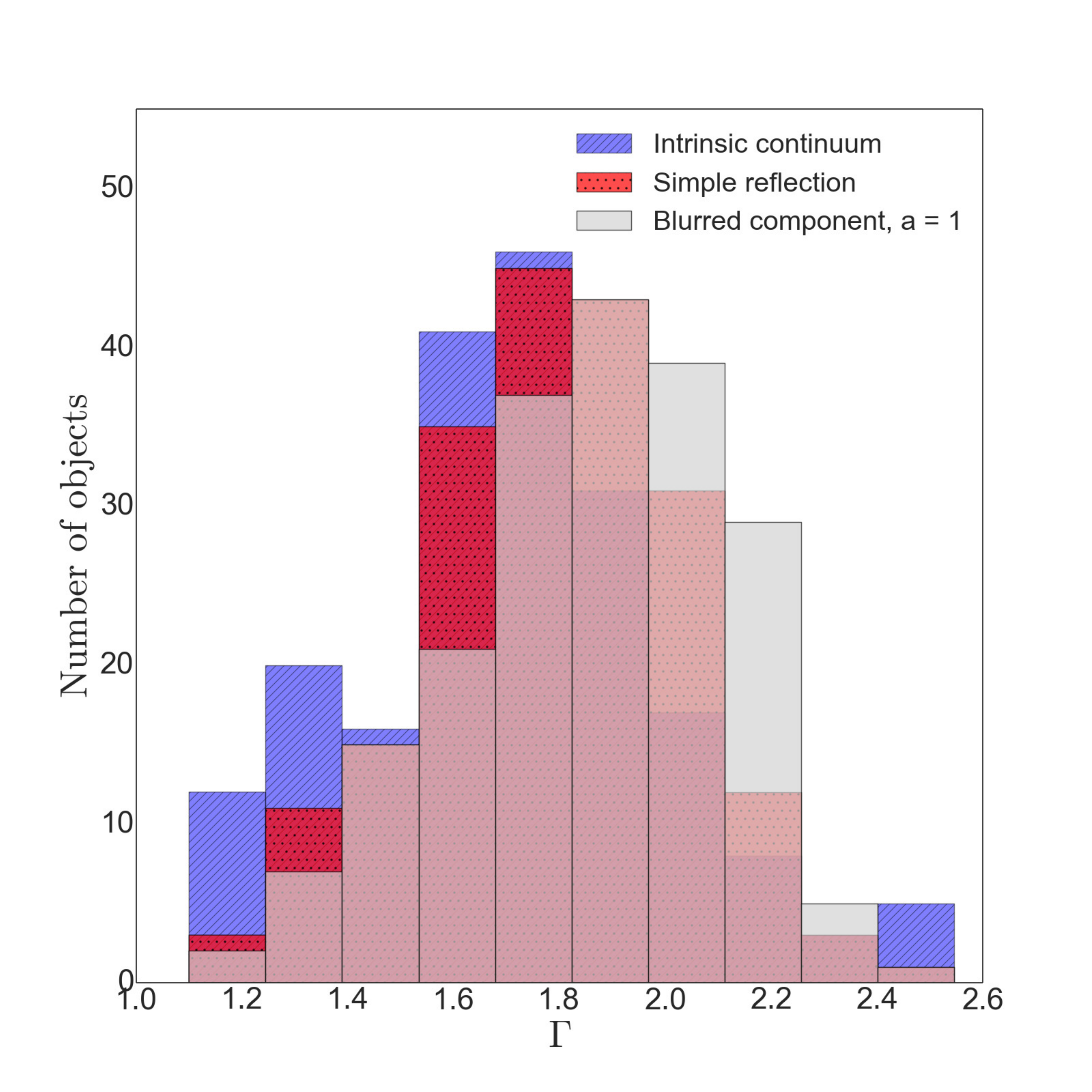}
     }
     \subfloat{%
       \includegraphics[width=0.35\textwidth]{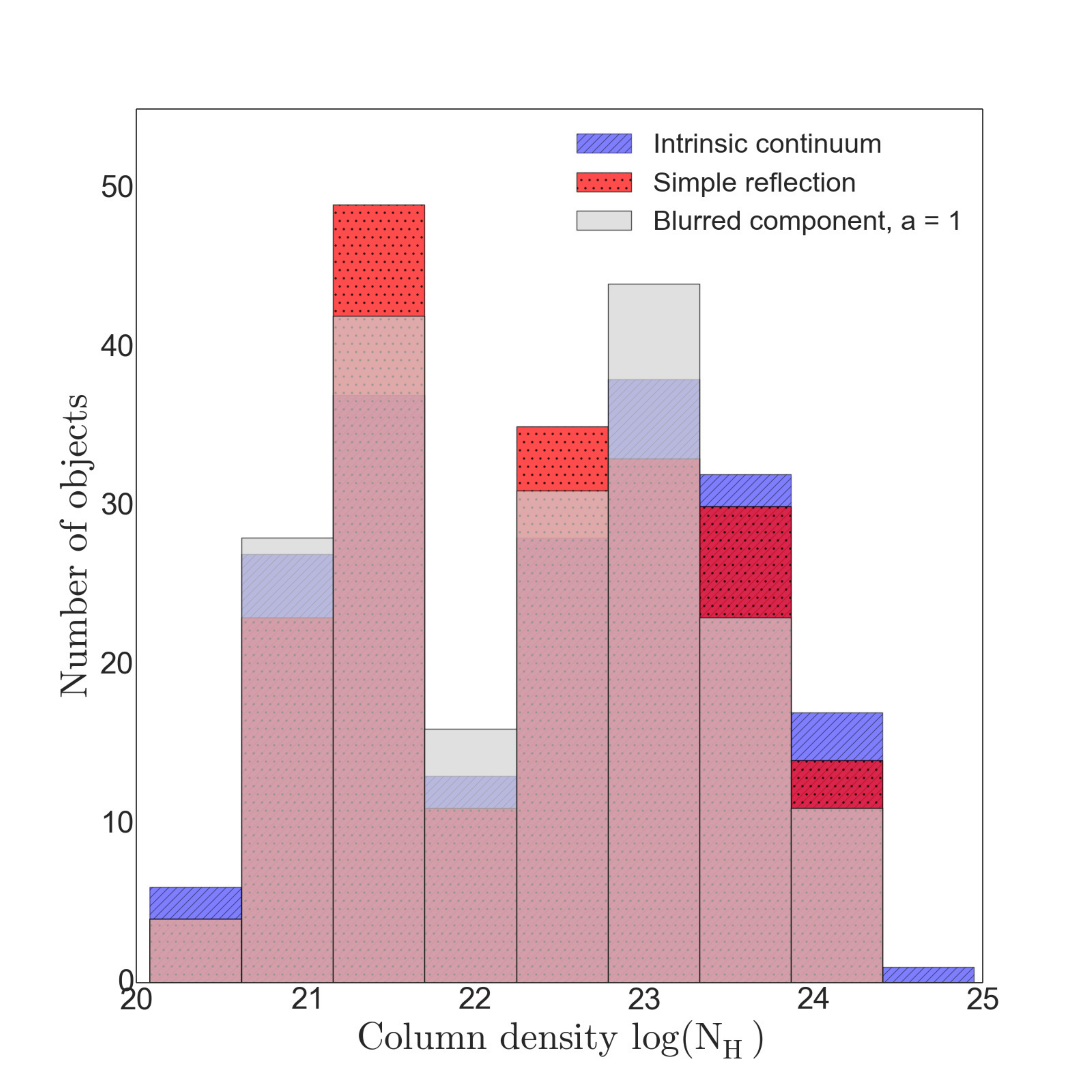}
     }
     \subfloat{%
       \includegraphics[width=0.35\textwidth]{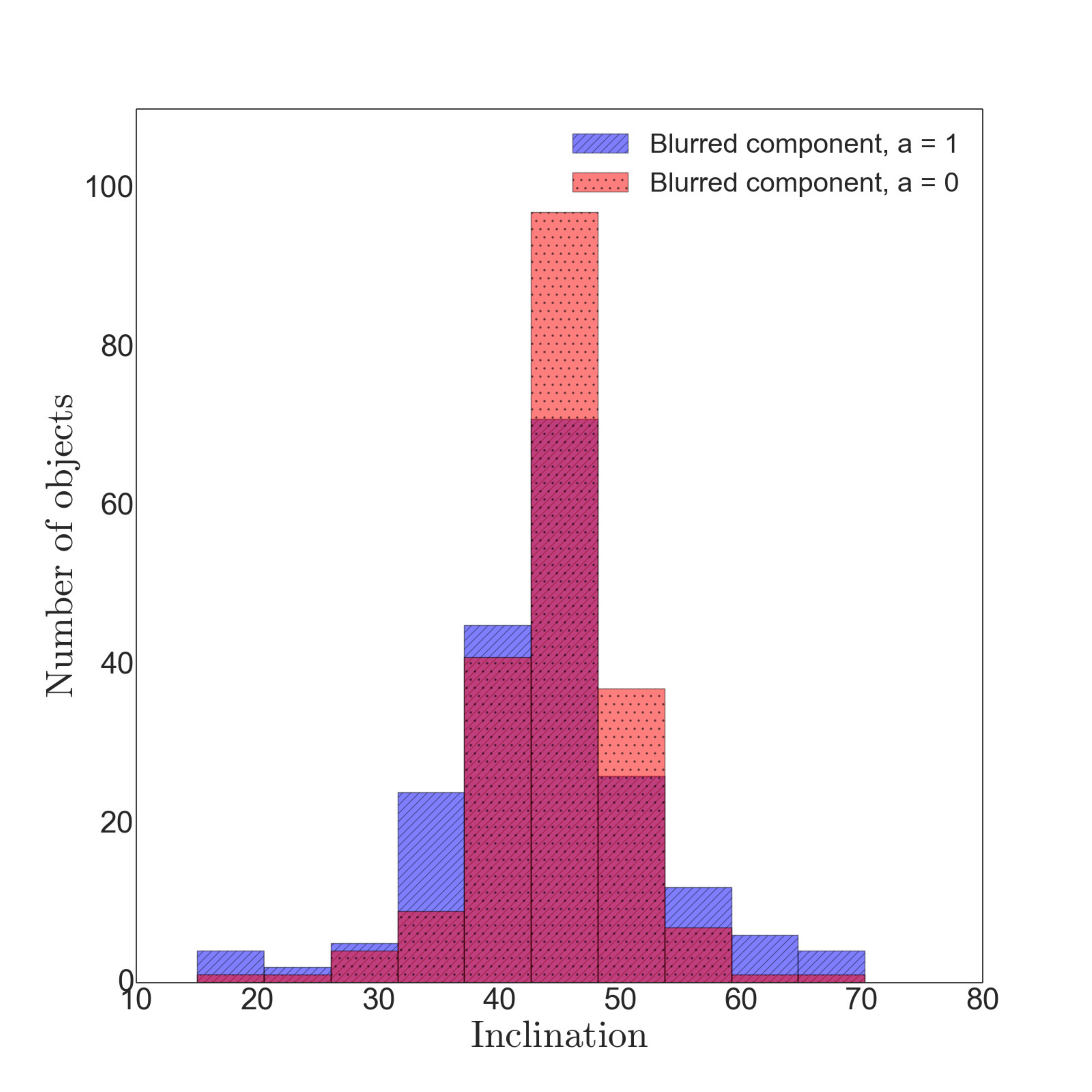}
     }
     \caption{\textit{Left}: Histogram of the best fit photon index $\mathrm{\Gamma}$ for the three models for the full sample. \textit{Middle}: Best fit column density $\mathrm{N_{H}}$ for the full sample. \textit{Right}: Comparison between the posterior distribution of the best fit inclinations for the blurred model with spin equal 1 (red, dotted) and the blurred model with spin 0 (blue, hatched).}
     \label{fig:all_param}
   \end{figure*}
\end{center}

% in the \textsc{sherpa} (\textsc{ciao} version 4.8) environment
We fit our sample of 199 sources using BXA, to determine the Bayesian evidence for the fitted models, the likelihood and the best fit parameters with their posterior probability distributions, presented in Figure \ref{fig:all_param}. %\textbf{In Figure \ref{fig:normBP} we report the R values of the narrow and broad reflection norms.} 
We use the Bayesian evidence to compare the models \ref{eq:powerlw}, \ref{eq:pexmon}, and \ref{eq:blur} and thus to determine which components are required to describe the data. In general, and as expected, the individual sources are usually too faint to select with high probability one model over another. However, we can add the Bayesian evidences of the single sources to determine the evidence for the whole sample. We find that the BF method \ref{eq:blur} selects the model containing a broad component as better fitting the full sample (see Table \ref{CDFSsample}) with respect to the model for narrow reflection \ref{eq:pexmon}. 
The blurred model is selected to be 10$^{5}$ times more probable than the narrow model (see Table \ref{CDFSsample}).
Furthermore, by comparing the blurred model with a=0 and the one with a=1, we find that the model representing a maximally spinning BH is selected by the BF method as 10$^{3.8}$ time more probable to be better fitting the data (see Table \ref{CDFSsample}).

\begin{figure*}
   \makebox[\textwidth][c]{
     \subfloat[CID 190.]{%
       \includegraphics[width=1.3\textwidth]{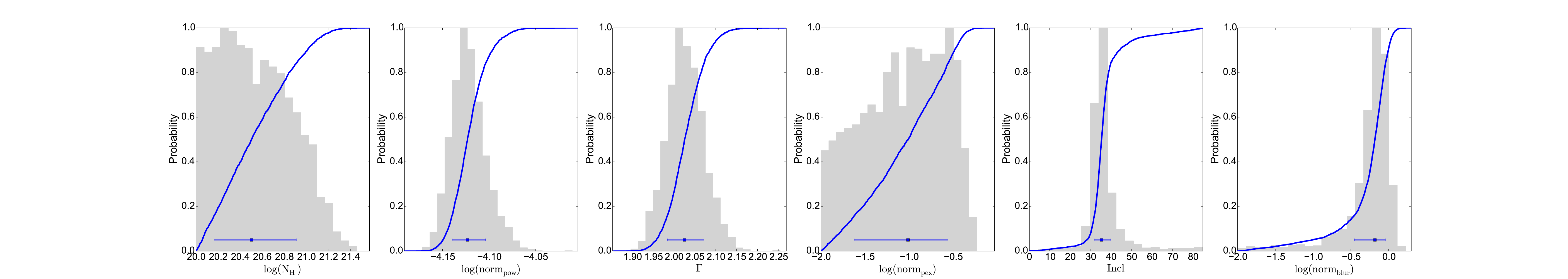}
     }}
     \\
     \makebox[\textwidth][c]{
     \subfloat[CID 104.]{%
       \includegraphics[width=1.3\textwidth]{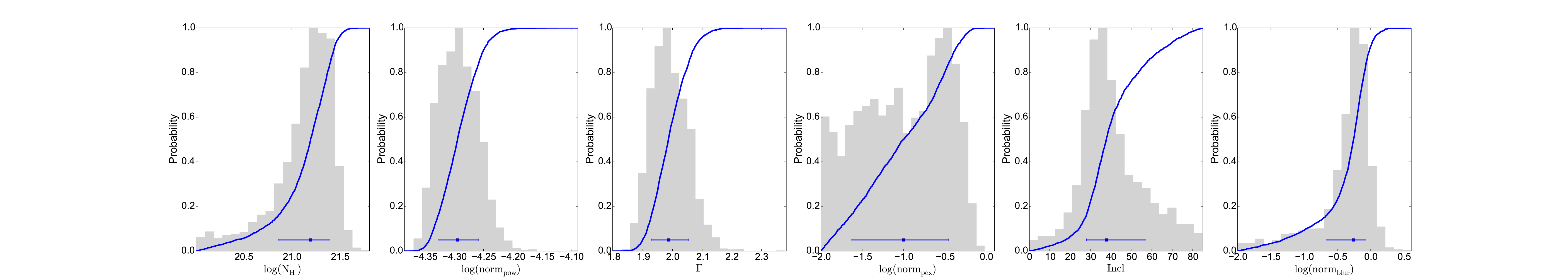}
     }}
     \caption{\textit{Top}: Marginalized parameters of the blurred model for CID 190. \textit{Bottom}: Marginalized parameters of the blurred model for CID 104. The marginals histogram is plotted in gray and the cumulative distribution function is over-plotted in blue. The inclination angle of the blurred component is $\sim 35^{+5}_{-4}$ degrees for CID 190, while the one of CID 104 is $\sim 37^{+20}_{-10}$ degrees. In both sources the normalization of the narrow component is not well constrained, as is also the case of the column density $\mathrm{log(N_H)}$ of CID 190}. The blue error bar indicates the 1 standard-deviation equivalent quantiles.
     \label{fig:marginals}
   
   \end{figure*}

\begin{center}

   \begin{figure}
     \subfloat{%
       \includegraphics[width=0.45\textwidth]{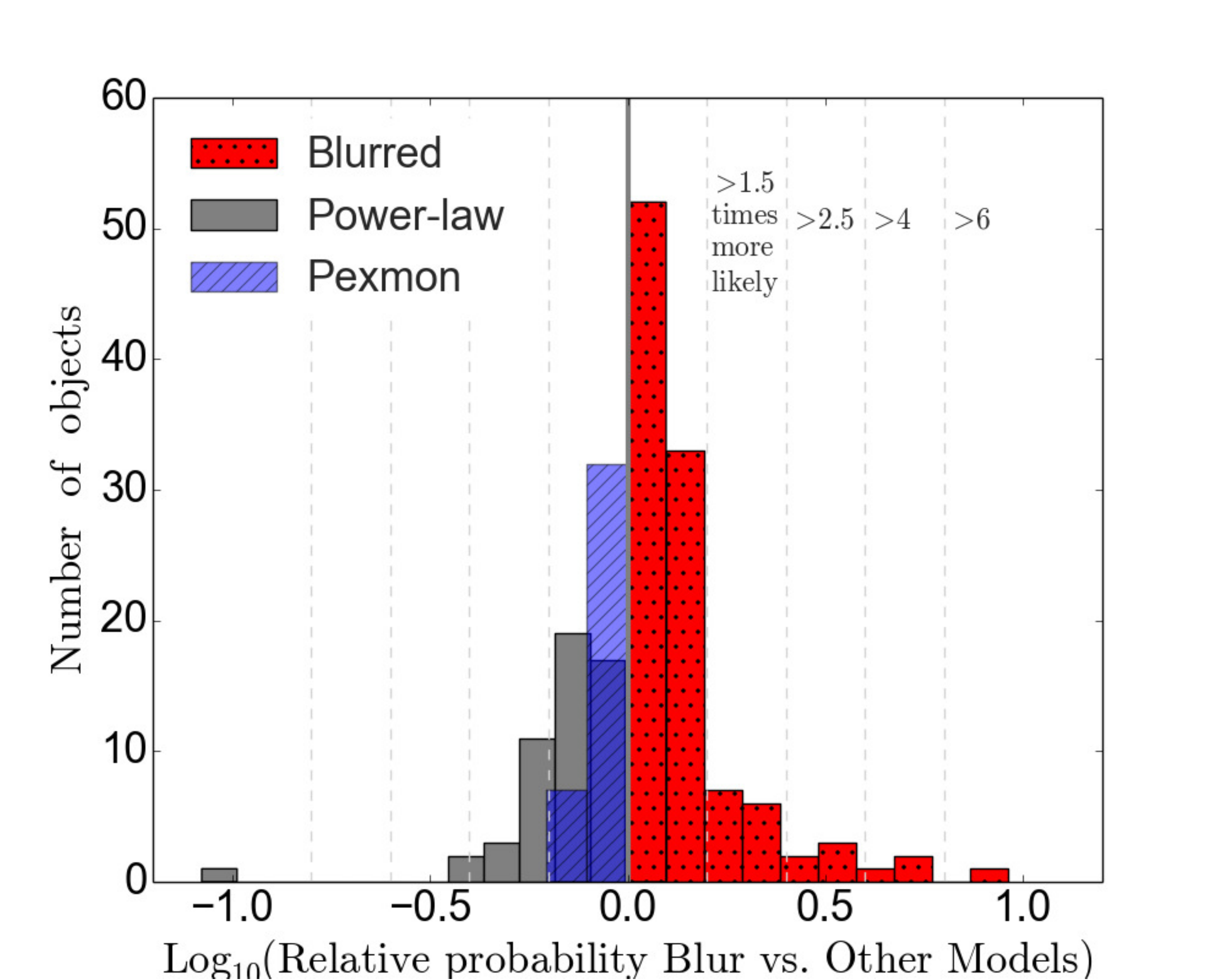}
     }
     \\
     \subfloat{%
       \includegraphics[width=0.45\textwidth]{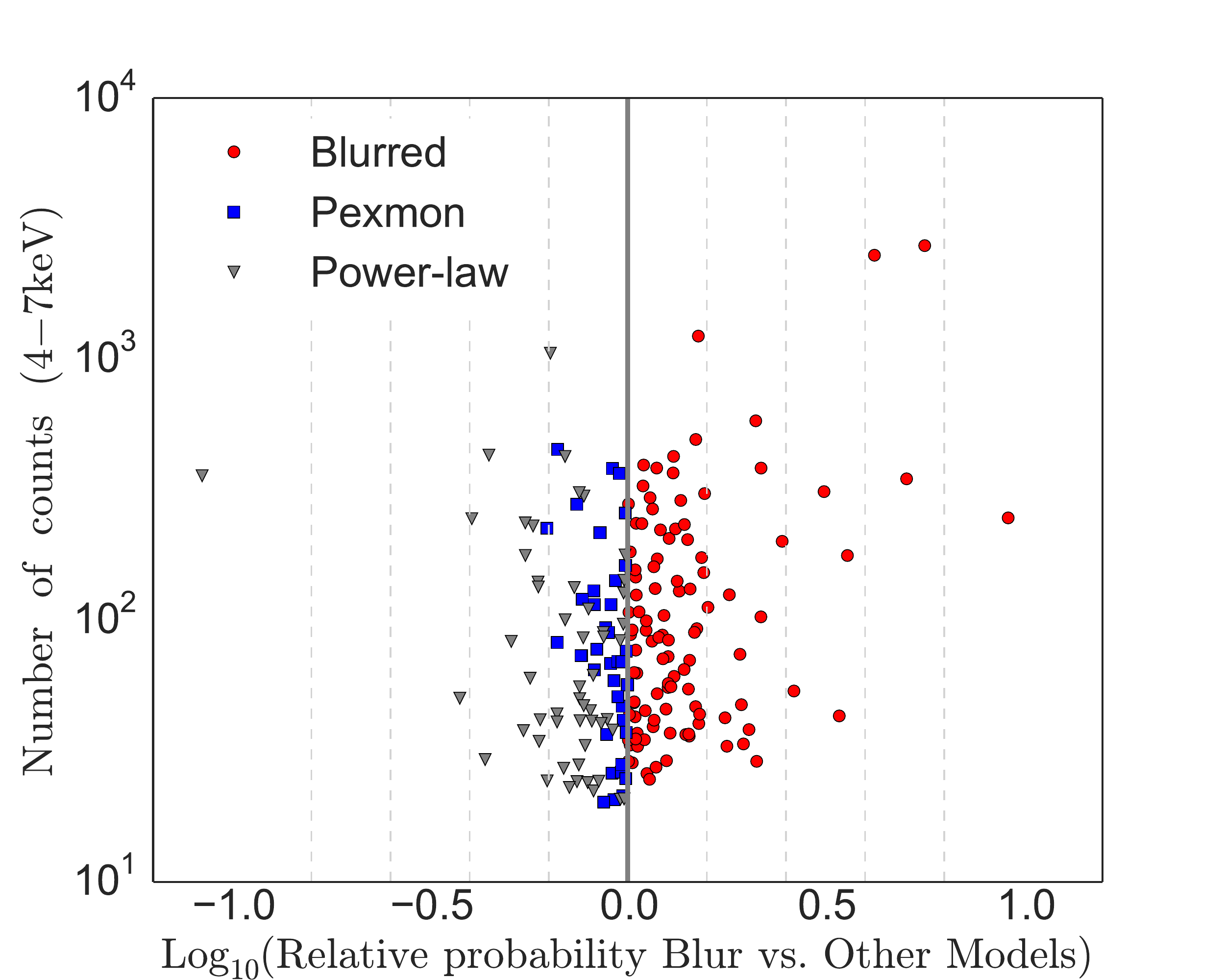}
     }
     \caption{\textit{Top}: Relative probability of the broadened model (spin 1) vs. the narrow reflection and the intrinsic emission. The sources are marked in red (dotted) is the model with highest probability is the blurred model \ref{eq:blur}, in blue hatched if the narrow model \ref{eq:pexmon} has the highest probability while in gray (solid) if model \ref{eq:powerlw} has the highest probability. \textit{Bottom}: Relative probability compared to the sources number of counts in the 4--7 keV. The single sources have such a low SNR that in the majority of the cases the blurred (circles) and the narrow (squares) models are equiprobable, even if the blurred one is slightly preferred. There are a few exceptions, e.g. CID 190 and CID 104, the two brightest sources of the sample. The sources with the power-law model having higher probability are marked as triangles.}
     \label{fig:rel_prob}
   \end{figure}
\end{center}

For the two brightest sources in our the sample, CID 190 ($z_{spec}=0.734$; ID numbers from R13) and CID 104 ($z_{spec}=0.543$) see Figure \ref{fig:weird}, the BF method also selects the blurred model with Kerr metric as best fitting the data. Using the BF method we obtain that for CID 190 model \ref{eq:blur} is twelve times more probable than model \ref{eq:pexmon}, thus the difference in Bayes evidence is $\mathrm{log_{10}\approx 1.1}$ which is 22$\%$ of the evidence difference between \ref{eq:blur} and \ref{eq:pexmon} of the total sample. Instead, for CID 104 model \ref{eq:blur} is five times as probable than model \ref{eq:pexmon}, hence its Bayes evidence difference ($\mathrm{log_{10}\approx 0.7}$) contributes to 13$\%$ of the difference in the evidence of the total sample. Hence, the contributions of CID 190 and CID 104 amount to the 35$\%$ of the total Bayesian evidence difference between the narrow and blurred models. Thus not taking the two brightest sources into account the Bayesian evidence difference would be of $\sim 10^{3.3}$ instead of $10^{5}$. CID 190 and CID 104 have respectively 2723 and 2502 counts in the 4--7 keV energy range and are the only sources in the sample with more than 2000 counts in that range. While they have large uncertainties, the inclination values we obtain for the blurred component in CID 190 and CID 104 are consistent with the inclinations obtained for other sources in the literature where the relativistic component of the Fe K$\alpha$ line was unambiguously observed. We obtain that the disk inclination of CID 190 is $\sim 35^{+5}_{-4}$ degrees, while the one of CID 104 is $\sim 37^{+20}_{-10}$ degrees (see Figure \ref{fig:marginals}).
While the vast majority of the individual sources have insufficient SNR to distinctly rule out one model over the others, we can still infer the fractions of sources containing a broad component by counting all the sources with highest Bayesian evidence for the broad model. 
Ranking the value of Bayesian evidence of the three models for each individual source, we find that the fraction of sources with highest evidence for the broad model, hence selected as containing the relativistic component is 54$\%$ (107/199). The sources selected as only containing a narrow reflection component are 19$\%$ (39/199), while the sources better described by a simple absorbed power-law comprise 27$\%$ (53/199) of the sample. These fractions have to be interpreted carefully, since in most cases the evidence difference between the three models is minimal (see Figure \ref{fig:rel_prob}).
%\textbf{ and here we do not apply any significant threshold to select one model over the others. To test the significance of the evidence difference for the single sources we constrain the numbers above to the sources with the highest evidence model at least twice as probable as the the other models. In this case, we find that only 14 sources have model \ref{eq:blur} twice as probable as the others models, 5 sources have model \ref{eq:powerlw} twice as probable and no source have the \ref{eq:pexmon} doubling the probability of the other models. The vast majority of the sources have very small differences in the evidence of the three models due to the low signal-to-noise of the spectra.} 
Nonetheless, the fraction of sources selected as presenting a relativistic broadened component obtained in this work is comparable with the fraction observed in \citet{Nandra2007} for local AGN.

\subsection{False positives and negatives}
\label{sec:false}

Because the difference in the evidence between the various models is generally small, statistical effects can result in both false positive detection for the relativistic components, or false negatives. To estimate the error on the selected fraction of sources showing a broad component we performed a set of simulations using the \texttt{fake$\_$pha} tool of \textsc{sherpa}. We simulated 200 sources for each of the spectral shapes given by models \ref{eq:powerlw}, \ref{eq:pexmon} and \ref{eq:blur}. 
The fake sources are simulated using the ancillary files of source CID 179 following the example of \citet{Buchner2014}, for each spectral shape we assign the power-law norm $5\times10^{-6}$ to the first 100 simulated sources and $10^{-5}$ to the remaining 100 sources. The strength of the blurred and narrow components with respect to the power-law norm is fixed to be $\mathrm{\log R = -0.3}$, which is the typical value found by \citep{Nandra2007}. This a conservative value, since if the actual R is smaller the number of false positives in the simulations will be overestimated. We fit the simulated sources using BXA to determine how many false positives and negatives we obtain by applying this method. The redshift of the simulated spectra is fixed at z=0.605, corresponding to the redshift of the original observation. 

By applying the three models to the spectra simulated using the \texttt{pexmon} model, we obtain that $\sim$63$\%$ of the simulated sources (126/200) are rightly selected as \texttt{pexmon} while $\sim$37$\%$ (74/200) are false positives, in that they are selected as containing a broad component even though we know that the underlying spectrum does not contain one. The total Bayes evidence shows correctly that the sample is better described by the \texttt{pexmon} model.

Similarly, if we fit the three analyzed models to the spectra simulated using a broad component we obtain that 65$\%$ (130/200) of the sources are correctly selected as broadened. 

The above analysis indicates that the inferred fraction of broadened components in our sample derived above ($\sim 54\%$) is likely to be a lower limit. For the typical signal-to-noise ratios in our sample around 35$\%$ of broad components would not be detected even if present, while 37$\%$ are false positives. If we consider only models \ref{eq:pexmon} and \ref{eq:blur} in the CDF-S sample without including model \ref{eq:powerlw} in the model comparison, we obtain that 63$\%$ (125/199) sources are selected as blurred while 37$\%$ (74/199) are selected as narrow. This result is very similar to the one obtained for the simulated sample with blurred component.
%Correcting for the false positive and negatives, we therefore infer crudely that the true fraction of sources displaying broad emission is approximately $\sim 67 \%$. 

\begin{table}
\caption{Comparison of the total sample Bayesian evidence for the models \texttt{zwabs*(zpowerlw)}, \texttt{zwabs*(zpowerlw+pexmon)} and \texttt{zwabs*(zpowerlw+pexmon+blur(pexmon))}. We fit the models in the observed frame energy range 1 -- 8 keV.}
\begin{tabular}{lr}
\hline \hline
Model $^{a}$ & \multicolumn{1}{l}{log10(Z)$^{b}$} \\ 
 \hline
\texttt{zwabs*(zpowerlw)} &  -78.1 \\ 
\texttt{zwabs*(zpowerlw+pexmon)} & -5  \\ 
\texttt{zwabs*(zpowerlw+pexmon+rdblur(pexmon))}& -3.8 \\ 
\texttt{zwabs*(zpowerlw+pexmon+kdblur(pexmon))}& 0 \\ 
 \hline
 \hline
\end{tabular}
\label{CDFSsample}
\flushleft 
\footnotesize{$^{a}$ Model components.}\\
\footnotesize{$^{b}$ Logarithm of the Bayes evidence of the full sample normalized to the largest evidence.}\\

\end{table}

\begin{center}

   \begin{figure*}
     \subfloat{%
       \includegraphics[width=0.55\textwidth]{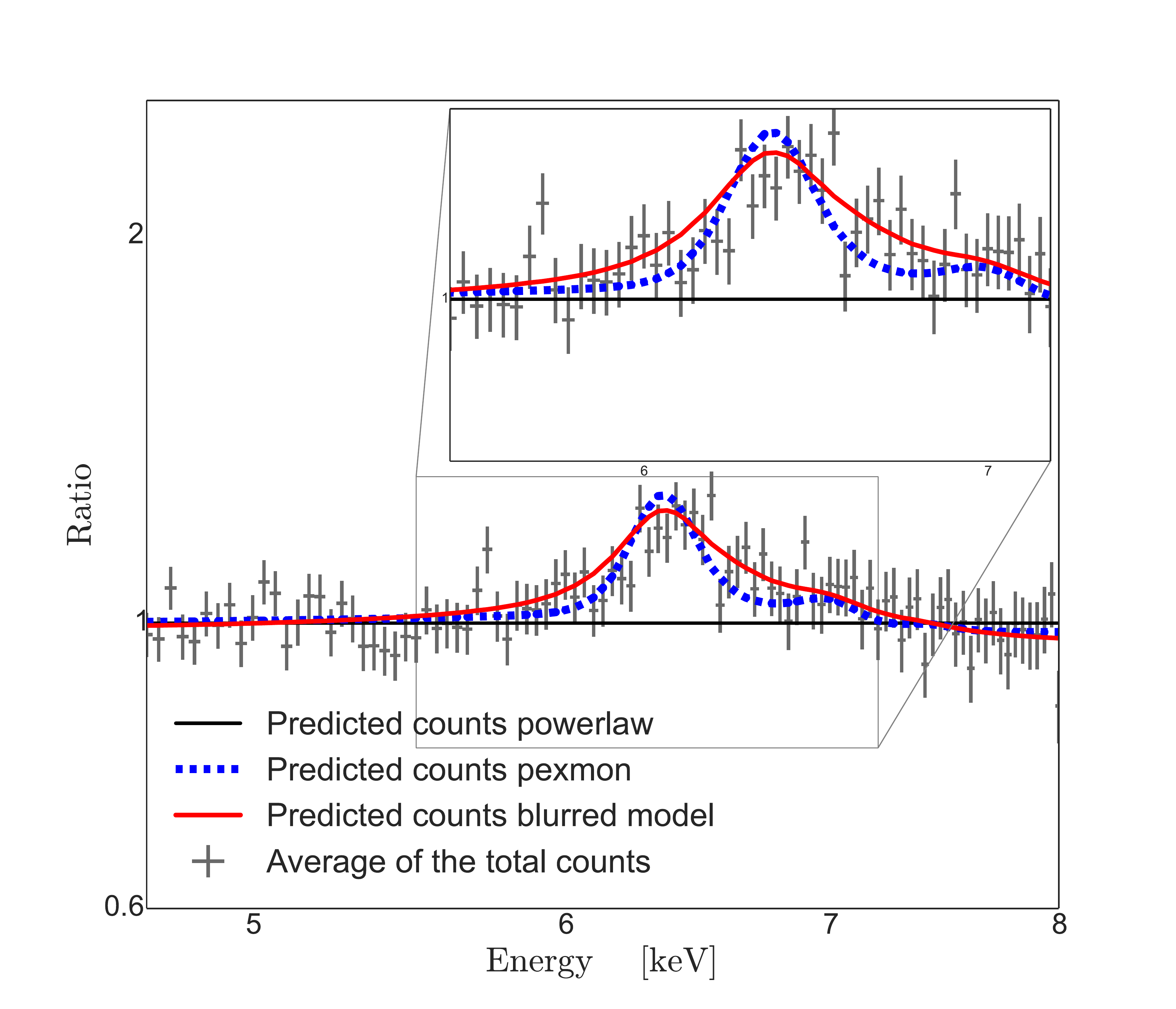}
     }
     
     \subfloat{%
       \includegraphics[width=0.55\textwidth]{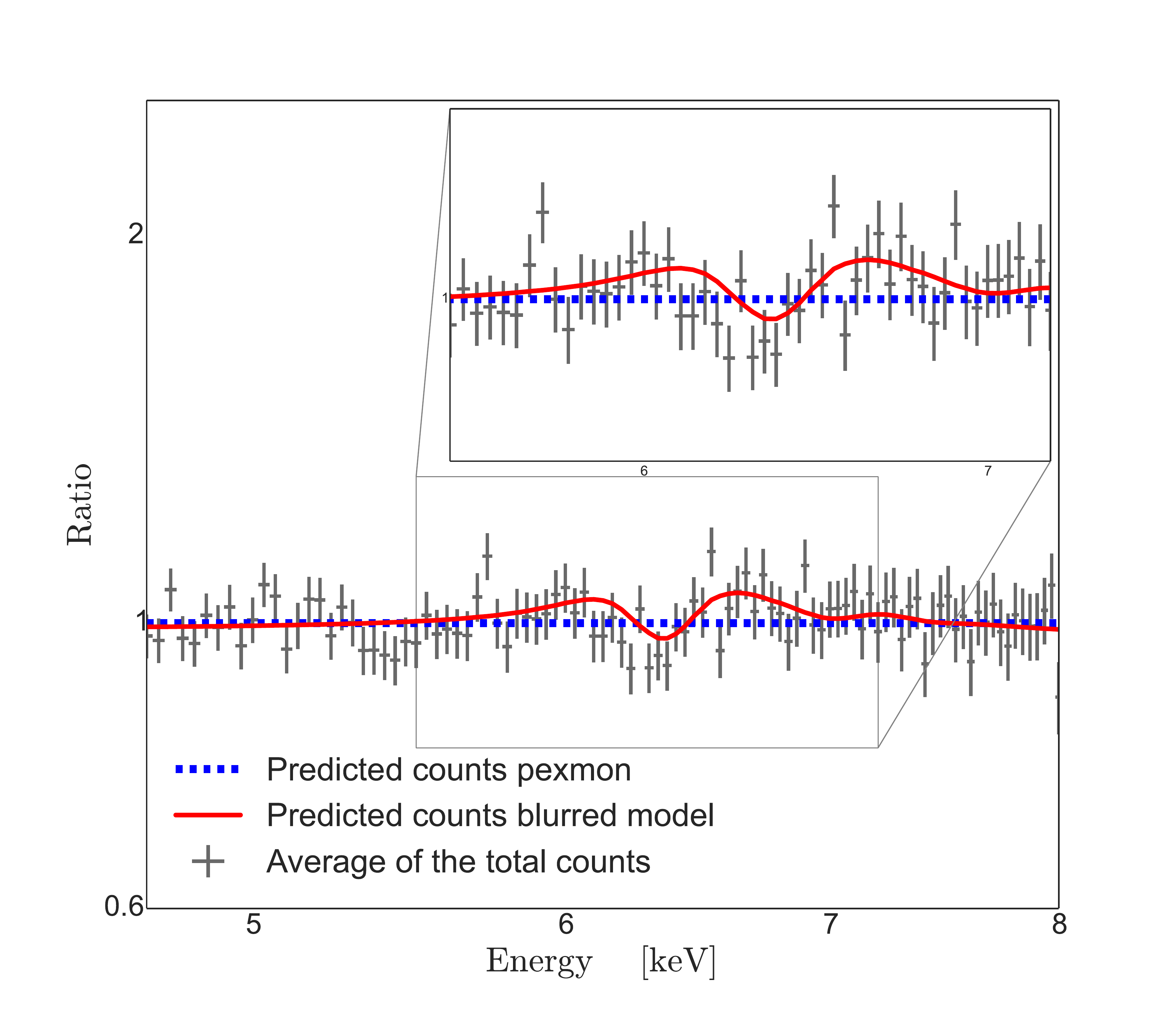}
     }
     \subfloat{%
       \includegraphics[width=0.55\textwidth]{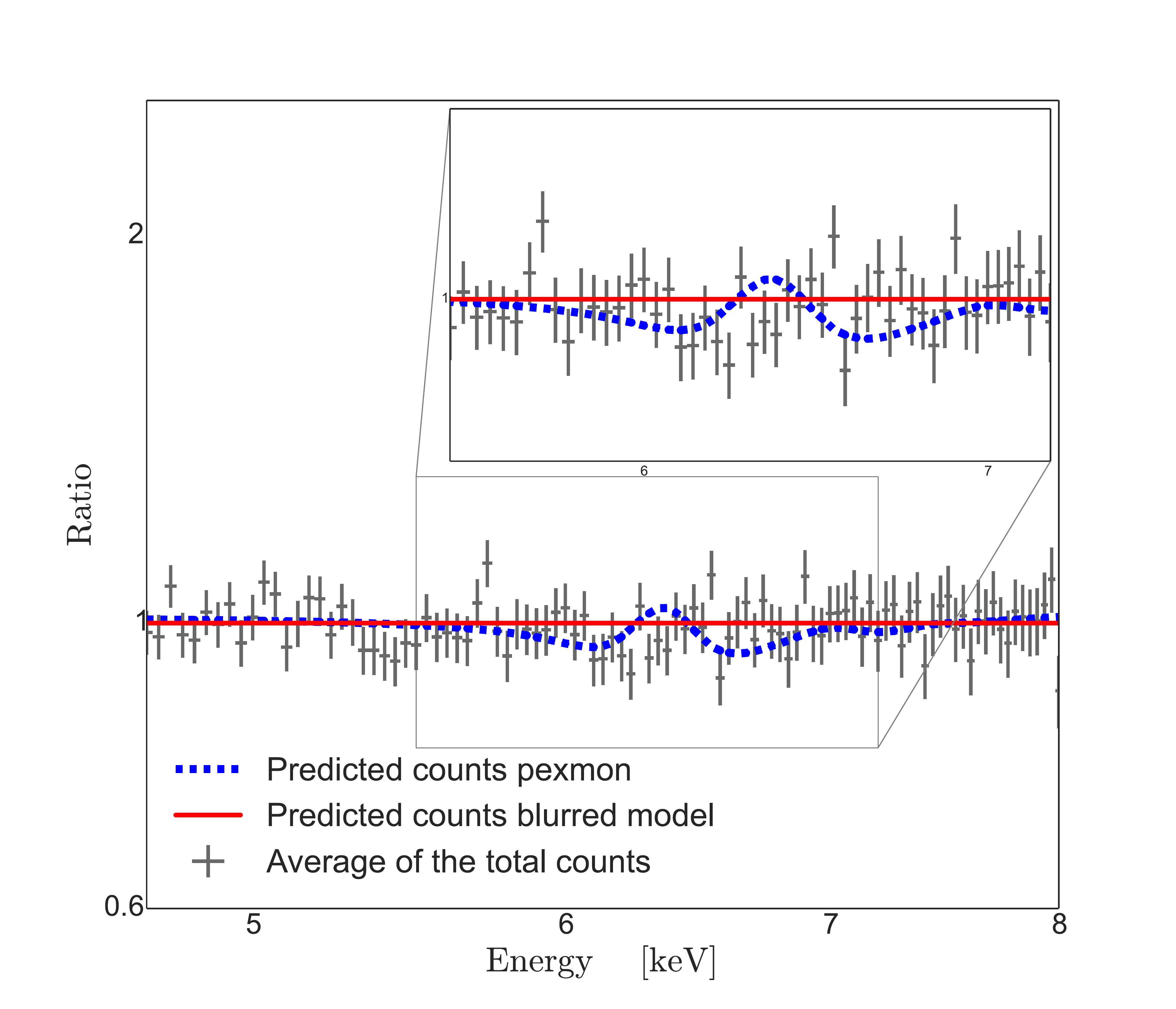}
     }
     \caption{\textit{Top}: Average of the total counts of the 199 spectra compared with the average of the three best fit models. The data and the models were normalized by the average of the best fitting power-law. \textit{Bottom left}: Same as above but in this case the data and the models are normalized by the narrow pexmon model (blue, dashed). \textit{Bottom right}: Same as above but data and models are normalized with the model including the blurred pexmon (red, solid). By comparing the two plots on the bottom it can be observed that the blurred model describes the data better than the narrow pexmon.}
     \label{fig:stack}
   \end{figure*}
\end{center}

\begin{comment}
\begin{figure*}
\includegraphics[scale=0.29]{ThreePan-eps-converted-to.pdf}
\caption{}
\label{fig:residuals}
\end{figure*}
\end{comment}

\begin{comment}
\begin{center}
\begin{figure}
\centering
\includegraphics[scale=0.28]{normBP-eps-converted-to.pdf}
\caption{Distribution of the narrow and broad reflection components for every source. In red we show the 1-$\sigma$ contour.}
\label{fig:normBP}
\end{figure}
\end{center}
\end{comment}

\section{Discussion}

The key finding of this work is evidence for the presence of a blurred Compton reflection component in the X-ray spectra of AGN in the CDF-S. This result has been established via fitting of the individual source spectra using their individual redshifts, and does not rely on spectral stacking, except for the purposes of visualization. This has been enabled by the use of the Bayesian spectral fitting procedure BXA, which allows us to perform parameter estimation and model comparison for our faint spectra, accounting for sources of error like statistical and systematic redshift uncertainty, Poisson uncertainty and background contribution. Previous evidence for broad emission has been found abundantly in the X-ray spectra of nearby Seyfert galaxies \citep[e.g.][]{Nandra2007} and has furthermore been claimed for fainter deep field AGN \cite[e.g.][]{Streblyanska2005,Chaudhary2012,Falocco2013,Falocco2014,Liu2016}, but in other cases has not been confirmed \citep[e.g.][]{Corral2008,Corral2011}. Our method, which fits individual spectra in the sample using a statistical method that is robust even with low counts, should provide a valid alternative to stacking.

The detection of relativistic emission in these sources is important, because it offers physical insight in to the accretion process in AGN at the peak of the cosmic accretion history at z=0.5--4 \citep[e.g.][]{Aird2010}. The accretion disk in these objects must extend very close to the supermassive black hole - indeed there is tentative evidence for emission within $6 R_{\rm g}$ based on the (mild) preference for a model with a rapidly spinning SMBH. Because the disk fluoresces at these radii in response to intense external illumination, it must therefore be relatively cool, in a relatively low ionization state, and hence of high density. All of this evidence points convincingly towards a standard, radiatively efficient accretion disk \citep[e.g.][]{Shakura1978} and argues strongly against hot, radiatively inefficient flows \citep[e.g.][]{Narayan1994,Hopkins2009}. 

Our results further suggest that these black holes may typically be rapidly spinning. While this still needs to be confirmed, it would suggest that these black holes grow via a fairly steady and long-term mode of accretion. Other modes such as "chaotic" accretion \citet{King2006} or black-hole black-hole mergers would tend to counteract the spin-up effect associated with long periods of coherent accretion. 

While it is certainly premature to consider these conclusions to be robust, they do indicate the tremendous power of X-ray spectroscopy to diagnose physical effects in these faint, growing supermassive black holes, and the future potential of this work. The fitting methods developed here rely not on the quality of the individual spectra, but on the total number of photons in the ensemble of spectra. They are therefore ideally suited to application to the forthcoming \textit{eROSITA} all-sky survey \citep{Predehl2010,Merloni2012}. The \textit{eROSITA} survey will yield AGN spectra with few counts, but for millions of X-ray emitting objects. Applying our techniques to these spectra will offer important insights into the relationship of the relativistic spectral signatures to other parameters such as accretion luminosity, accretion rate, black hole mass, obscuration and perhaps even larger-scale galaxy properties or large-scale structure environment. A revolution in such studies will later be provided by \textit{Athena} \citep{Nandra2013}, whose unprecedented collecting area will have the potential to reveal relativistic signatures in individual cases, as well as ensemble samples such as those shown here. 

\section{Summary and Conclusions}
\label{sec:Summary_Conclusion}

Using a sample of 123 X-ray observations of AGN with spectroscopic redshift and 76 with photometric redshift 
from the CDF-S, we have sought to establish whether a relativistic broadened Fe K$\alpha$ line and Compton reflecting continuum is a common characteristic of X-ray AGN detected in the CDF-S. 
This has been achieved by fitting the spectra individually, rather than stacking, and then selecting the best fit model via Bayesian model comparison. Our main findings are: 
\begin{itemize}
\item The Bayesian evidence of the full sample shows that the model containing a relativistically blurred component is preferred over those without such a component. 

\item  The data show a preference for a spinning SMBH, specifically a Schwarzschild SMBH with spin a=0 (\texttt{rdblur}) fits the data less well than one with a Kerr SMBH with a=1 (\texttt{kdblur}). 

\item Observations of the the two brightest sources in the sample (CID 190 and CID 104) confirm the results for the sample as a whole, in that the blurred model with a high spin is preferred when fitting their spectra individually.

\item The estimated the fraction of objects showing a blurred component is approximately 63$\%$, but this can be considered a lower limit given the low-signal-to-noise ratio of the spectra and the penalization of the more complex blurred model without a disk component. 

\item Our results imply that the majority of black hole growth in the Universe proceeds via standard, radiatively disk accretion, and demonstrate the great future potential of X-ray spectroscopy to reveal the physics of accretion, out into the high redshift Universe. 
\end{itemize}

\newpage

%-----------------------------------------------------------------------------------------------------------------------------------

\section*{Acknowledgements}

This research made use of {\tt Astropy}, a community-developed core Python package for Astronomy (Astropy Collaboration, 2013) and the NASA's Astrophysics Data System. This research made use of APLpy, an open-source plotting package for Python \citep{Robitaille2012}. We also used extensively the Python package Matplotlib \citep{Hunter2007}.

JB acknowledges support from the CONICYT-Chile grants Basal-CATA PFB-06/2007, FONDECYT Postdoctorados 3160439 and the Ministry of Economy, Development, and Tourism's Millennium Science Initiative through grant IC120009, awarded to The Millennium Institute of Astrophysics, MAS.

%%%%%%%%%%%%%%%%%%%%%%%%%%%%%%%%%%%%%%%%%%%%%%%%%%


\bibliographystyle{mnras}
\bibliography{Broad_CDF-S} % if your bibtex file is called example.bib

%%%%%%%%%%%%%%%%%%%%%%%%%%%%%%%%%%%%%%%%%%%%%%%%%%

% Don't change these lines
\bsp	% typesetting comment
\label{lastpage}
\end{document}